\documentclass{acm_proc_article-sp}
\usepackage{booktabs}
\usepackage{times}
\usepackage{graphicx}
\usepackage[hyphens]{url}
\usepackage{hyperref}
\usepackage{arabtex}
\novocalize

\begin{document}

\title{The Hush Cryptosystem}

\numberofauthors{1}

\author{
\alignauthor Sari Haj Hussein\\
       \affaddr{Department of Computer Science\\Chalmers University of Technology\\Gothenburg, Sweden}\\
       \email{angyjoe@gmail.com}
}

\maketitle

\begin{abstract}
In this paper we describe a new cryptosystem we call "The Hush Cryptosystem" for hiding encrypted data in innocent Arabic sentences. The main purpose of this cryptosystem is to fool observer-supporting software into thinking that the encrypted data is not encrypted at all. We employ a modified Word Substitution Method known as the Grammatical Substitution Method in our cryptosystem. We also make use of Hidden Markov Models. We test our cryptosystem using a computer program written in the Java Programming Language. Finally, we test the output of our cryptosystem using statistical tests.
\end{abstract}




\section{Introduction}\label{Introduction}
The great popularity of information systems and the expansion in size and speed in data transfer operations make it no longer possible for those who want to monitor a communication channel to rely on human factors solo. Instead, it is necessary to use software that analyzes the data, and chooses a sample of interest for the observer to review. Most of the data transferred through communication channels are unencrypted, something the observer is uninterested in as he expects the secret data to be sent in an encrypted form. Thus, the main purpose of the observer-supporting software is to recognize encrypted data, filter it, and then provide it to the observer.
After having the encrypted data, the observer will conduct a cryptanalysis operation in order to decrypt the data. The operation may be unsuccessful at first when having few data; however, and as the amount of data grows, the observer may be able to conduct statistical tests that would undoubtedly raise the probability of a successful cryptanalysis. We can prevent the observer from having the encrypted data by hiding it as plain data. In that case, the observer-supporting software will not be able to detect encrypted data and provide it to the observer. This is actually the main purpose of this study. In this paper, we suggest replacing random data that results from an encryption operation with words from a natural language so that the randomness degree of encrypted data decreases to equate that of plain data. How do we achieve our goal? We need to know the tools available in observe-supporting software. The first tool relies on conducting statistical tests to determine how random the data is and then tell whether it is encrypted or not; eventually, the main characteristic of encrypted data is that it is more random than plain data. The second tool available in observer-supporting software relies on testing how close the structure of the data is to that of natural languages. Using randomly scattered words is never enough. Instead, sentences should be grammatically correct. The last tool relies on verifying that sentences are meaningful. A system for Arabic language understanding and detecting spurious Arabic sentences never exits until now; we will get back to this issue in Section \ref{English-Why-Not}. Thus, we can say that replacing encrypted data with grammatically correct Arabic sentences suffices to bypass the observer-supporting software.

\section{Natural Language Words}\label{Natural-Language-Words}
The first step in hiding encrypted data is converting it into natural language letters and words in order to decrease the Randomness Degree. There exist renowned methods for this. For us to evaluate these methods, we need to take two considerations into account:
\begin{enumerate}
  \item Randomness Degree: Every method has an associated Randomness Degree. A method may achieve a Randomness Degree that is too close to that of natural language texts; but it may fail at satisfying the second consideration which is:
  \item Redundant Data Percentage: The more the redundant data in encryption output, the higher the probability of decryption. Therefore, the Redundant Data Percentage should not exceed a certain upper bound.
\end{enumerate}
We will discuss three frequently used methods for converting random data to natural language letters and words. They are Base64 Encoding, Sentence Substitution and Word Substitution.

\subsection{Base64 Encoding}\label{Base64-Encoding}
This encoding converts binary data into data formed of ASCII characters only \cite{RFC}. The main purpose of this encoding is to facilitate data transfer through communication channels that allow ASCII characters only such as e-mail channels. Also it may be used for hiding encrypted data in natural language letters. This encoding is widely used because of it offers great balancing between simplicity and efficiency. The Base64 Encoding converts every $3$ bytes to $4$ ASCII characters. That is, every byte is represented by $6$ bits only. Therefore, we need $2^{6}=64$ character in Base64 Encoding.

\subsubsection{Base64 Randomness}\label{Base64-Randomness}
The Base64 Randomness Degree decreases as a result of the decrease in the number of characters from $256$ (in plain data) to $64$ in Base64; however, this decrease is never enough to bypass randomness tests. In addition, this encoding generates meaningless characters that can never masquerade as plain data before observer-supporting software.

\subsubsection{Base64 Redundancy}\label{Base64-Redundancy}
Every $6$ bytes are converted to $8$, so the redundancy is $2$ bytes, and the percentage is $33.3\%$ of the original data which is satisfying.

\subsection{Sentence Substitution}\label{Sentence-Substitution}
The concept behind this method is simple \cite{Stamp}. It works by replacing every group of bits with a natural language sentence. The number of bits $n$ in every group depends on the number of available sentences $m$. We create a table that matches every group of bits with a sentence. The number of bits and sentences satisfies the formula:
\begin{equation*}
    m\geq2^{n}
\end{equation*}
This method generates a series of natural language sentences that may seem logically disjointed to humans; however, observer-supp-\\orting software will not be able to distinguish between them and between plain data.

\subsubsection{Sentence Substitution Randomness}\label{Sentence-Substitution-Randomness}
As long as we are replacing with natural language sentences, the Randomness Degree of the resulting text is close to that of plain data. Thus the first consideration of Section \ref{Natural-Language-Words} is satisfied.

\subsubsection{Sentence Substitution Redundancy}\label{Sentence-Substitution-Redundancy}
This method results in a big increase in the amount of data because we are replacing a small number of bits with a complete sentence of multiple bytes. Assume, for example, that we have $1024$ sentences, and that we are encoding every $10$ bits with a sentence, and that the length of the sentence is $25$ letters (on average). Then to encode $8$ bytes ($64$ bits), we need $7$ sentences or $175$ letters. That is a Redundant Data Percentage of $97\%$ which is not satisfying at all. Another disadvantage of this method is that sentences will be eventually repeated when encoding a fairly large amount of data. If that happened, it would be very easy to discover the method of hiding and then reclaim the original data.

\subsection{Word Substitution}\label{Word-Substitution}
We said that the problem with Sentence Substitution is the unbearable redundancy in the resulting text. The next idea we propose is to replace every group of bits with a natural language word taken from a dictionary \cite{Chapman1997}. This allows us to control data increase. The larger the dictionary used, the larger the group of bits that can be replaced with one word. That is, the larger the dictionary, the smaller the redundancy. As with Sentence Substitution, the following formula, between the number of bits $n$ in every group and the number of words in the dictionary $m$, holds:
\begin{equation*}
    m\geq2^{n}
\end{equation*}
If we assumed, for example, a dictionary of $131072$ words, then we can encode every $\log_{2}131072=17$ bits (i.e. every $2$ bytes approximately) with one word. We can also utilize the characteristics of natural languages to increase the number of available words. In Arabic, for example, we can add prefixes and suffixes to obtain a variety of new words. Let us consider an example to illustrate this method further. Assume an Arabic dictionary of two words only: \RL{aby.d} (Arabic for white), and \RL{aswd} (Arabic for black). Then, every single bit is encoded with one of theses two words. Assume further that \RL{aby.d} replaces $0$, and \RL{aswd} replaces $1$, then the byte $10110110$ is encoded as shown in Table \ref{Encoding-the-byte}. Or:
\begin{equation*}
    \RL{aswd}\;\RL{aby.d}\;\RL{aswd}\;\RL{aswd}\;\RL{aby.d}\;\RL{aswd}\;\RL{aswd}\;\RL{aby.d}
\end{equation*}
This example demonstrates that recognizing the hiding method is rather easy when using Word Substitution; however, when a large dictionary is used, the recognition operation becomes more challenging.

\begin{table}
    \centering
    \caption{Encoding the byte $10110110$ using a two-word Arabic language dictionary}
    \label{Encoding-the-byte}
    \begin{tabular}{llllllll}
    \midrule
    $1$ & $0$ & $1$ & $1$ & $0$ & $1$ & $1$ & $0$\\
    \RL{aswd} & \RL{aby.d} & \RL{aswd} & \RL{aswd} & \RL{aby.d} & \RL{aswd} & \RL{aswd} & \RL{aby.d}\\
    \bottomrule
    \end{tabular}
\end{table}

\subsubsection{Word Substitution Randomness}\label{Word-Substitution-Randomness}
Because it is based on using natural language words, the Word Substitution Method achieves a Randomness Degree that is too close to that of plain data, thus observer-supporting software will not be able to distinguish between hidden data and plain data using statistical tests alone.

\subsubsection{Word Substitution Redundancy}\label{Word-Substitution-Redundancy}
The Redundant Data Percentage in Word Substitution is far less than it is in Sentence Substitution. Instead of replacing a group of bits with a complete sentence, we are replacing with one word. Assume, for example, a dictionary of $131072$ words, and that we replace every $17$ bits with one word. If we assumed further that the average word length is $5$ letters, we find that the Redundant Data Percentage is $57.5\%$ which is much better than it is in Sentence Substitution. This decrease in Redundant Data Percentage comes with a price to pay. The price is that the generated texts are meaningless for humans. But as long as the purpose of this study is to bypass observer-supporting software, and not to bypass humans! We realize that generating meaningful Arabic texts is unimportant because software for Arabic language understanding never exists until now. What software can verify for now is the syntactic validity of sentences structures. Therefore, we will develop the method of Word Substitution to overcome this problem.

\subsubsection{Word Substitution Implementation}\label{Word-Substitution-Implementation}
The implementation of Word Substitution is fairly simple, and could be summarized in the following:
\begin{enumerate}
    \item Determine the number of words $m$ in the dictionary and calculate the number of bits $n$ in every group using the formula:
    \begin{equation*}
        n=\mathit{floor}(\log_{2}m)
    \end{equation*}
    \item Number the words in the dictionary from $0$ to $m-1$.
    \item Set up a function that matches the decimal value of every group of bits with a word number in the dictionary. If we denoted the decimal value of the group of bits as $p$ and the number of the corresponding word as $q$, then we can set up the following function:
    \begin{equation*}
        f:\mathit{IN}_{2^{n}}\rightarrow \mathit{IN}_{m}:q=f(p)
    \end{equation*}
    whereas the inverse function is:
    \begin{equation*}
        g=f^{-1}:\mathit{IN}_{m}\rightarrow \mathit{IN}_{2^{n}}:p=g(q)    
    \end{equation*}
    \item Divide the original data into groups of $n$ bits, calculate the decimal value of the bits $p$, and then apply the function $f$ to obtain the number of the corresponding word $q$.
\end{enumerate}
The reverse operation for obtaining original data from the resulting text is as follow:
\begin{enumerate}
    \item Search for the word in the dictionary and determine its number $q$.
    \item Apply the inverse function $g$ to obtain the decimal value of the group of bits $p$.
    \item Convert $p$ into the binary system to get the original set of data.
\end{enumerate}
The function $f$ may be based on the concept of public and private keys, or simply on the concept of the Identity Function $I$ if the original data were strongly encrypted.

\section{Natural Language Grammar}\label{Natural-Language-Grammar}
In order to overcome the problem with Word Substitution, we will generate sentences with a structure similar to that of natural language sentences. Our theoretical study will be made as general as possible, but the practical examples will be given for Arabic. We will use Hidden Markov Models, which are commonly used in the processing of natural languages.

\subsection{Hidden Markov Models}\label{Hidden-Markov-Models}
\begin{figure}
  \centering
  \includegraphics[scale=0.15]{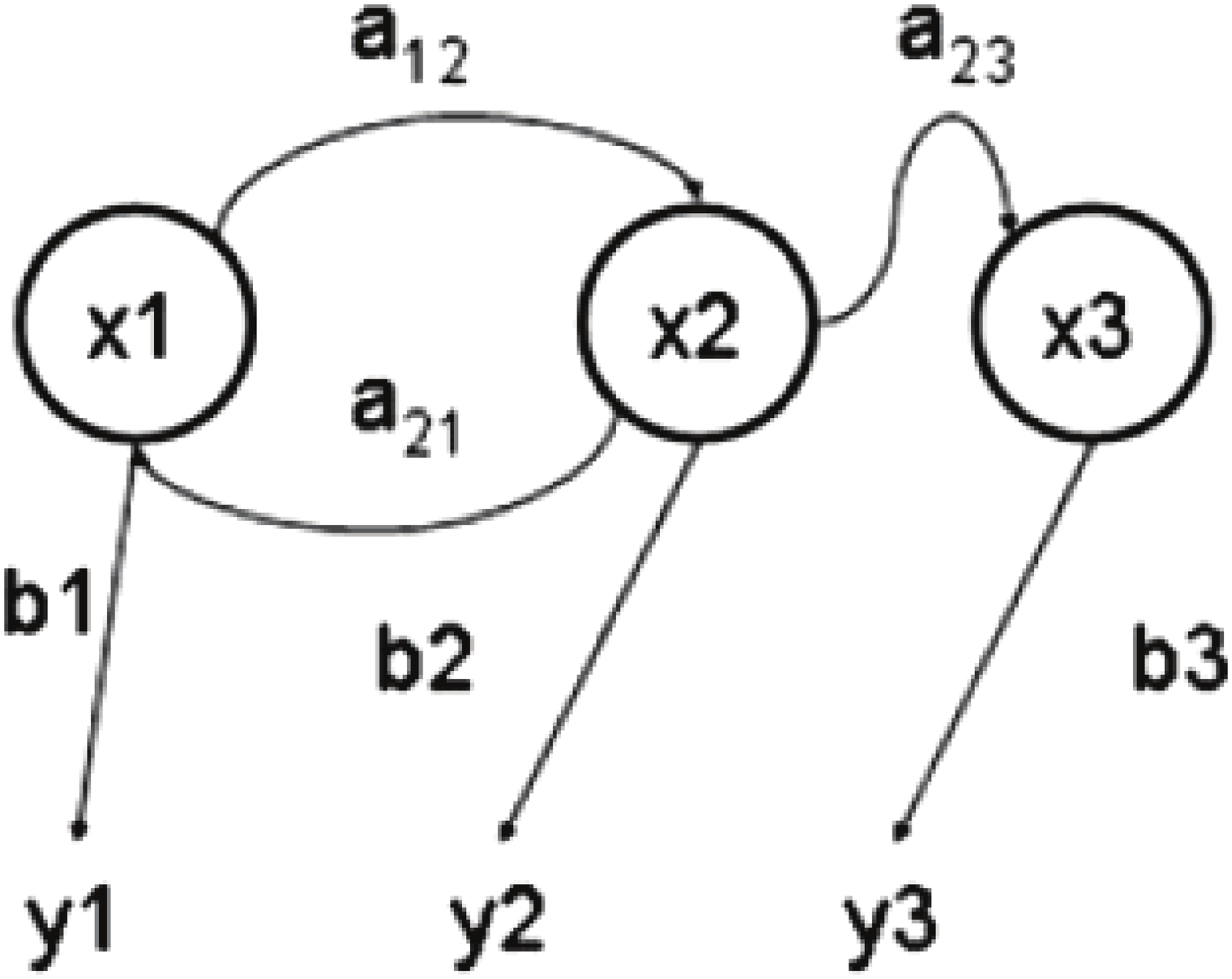}
  \caption{A Hidden Markov Model}\label{A-Hidden-Markov-Model}
\end{figure}

A Hidden Markov Series \cite{Rabiner1990} is a statistical model in which the system is a Markov Series with unknown parameters. The output of this system is the only thing available for us, and we have to use it to determine the values of the unknown parameters. In a regular Markov Model, the state of the system is known to those who study the system, and the probabilities of transition are the only parameters in the system. A Hidden Markov Series adds an output to the system where every state has a probability of generating a specific output. Thus, we can not determine the states of the system by just looking at the output. And that is why these Markov Models are called "Hidden". Hidden Markov Models are used in the study of systems that generate probabilistic patterns where a natural language can be seen as one such pattern. A Hidden Markov Model is illustrated in Figure \ref{A-Hidden-Markov-Model} where $x_{i}$: the State $i$, $a_{ij}$: the probability of transition from state $i$ to state $j$, $y_{i}$: the output $i$, and $b_{i}$: the probability of generating output $i$.

\subsubsection{Deterministic Patterns}\label{Deterministic-Patterns}
A deterministic pattern has specific states, and the transition of the system to a new state depends only on its current state regardless of the previous states or any other values. These systems are similar somehow to road signs as road signs have only three states; red, orange, and green, and the transition of the system to a new state (orange for example) depends only on its current state (red).

\subsubsection{Probabilistic Patterns}\label{Probabilistic-Patterns}
Deterministic patterns are never enough to represent all systems. The one who is studying the system may not have the rules for system transition from one state to another. He may only have the results of system transition between its states. Thus, he will have to determine the states and the rules using these results. Natural languages are examples on this. When we analyze a text, we see the results of applying the grammars of a natural language; these results are the sentences that form the text. However; these grammars are not readily available for us. To get them, we must study the output of the system. We see here the advantage of Hidden Markov Models in forming sentences that follow the grammars of a natural language.

\subsubsection{Using Hidden Markov Models}\label{Using-Hidden-Markov-Models}
There are three primary problems with Markov Models; problems when solved, enable us to solve any other problem related to these models. The three problems are:
\begin{enumerate}
    \item Given system parameters, we want to calculate the probability of generating a specific output. This problem is solvable using the Forward Algorithm.
    \item Given system parameters, we want to find the pattern of hidden states that generated a specific output. This problem is solvable using the Viterbi Algorithm.
    \item Given some output, we want to find the pattern of hidden states and the highest probability output. This problem is solvable using the Baum-Welch Algorithm.
\end{enumerate}

\subsubsection{Example}\label{Example}
\begin{figure}
  \centering
  \includegraphics[scale=0.15]{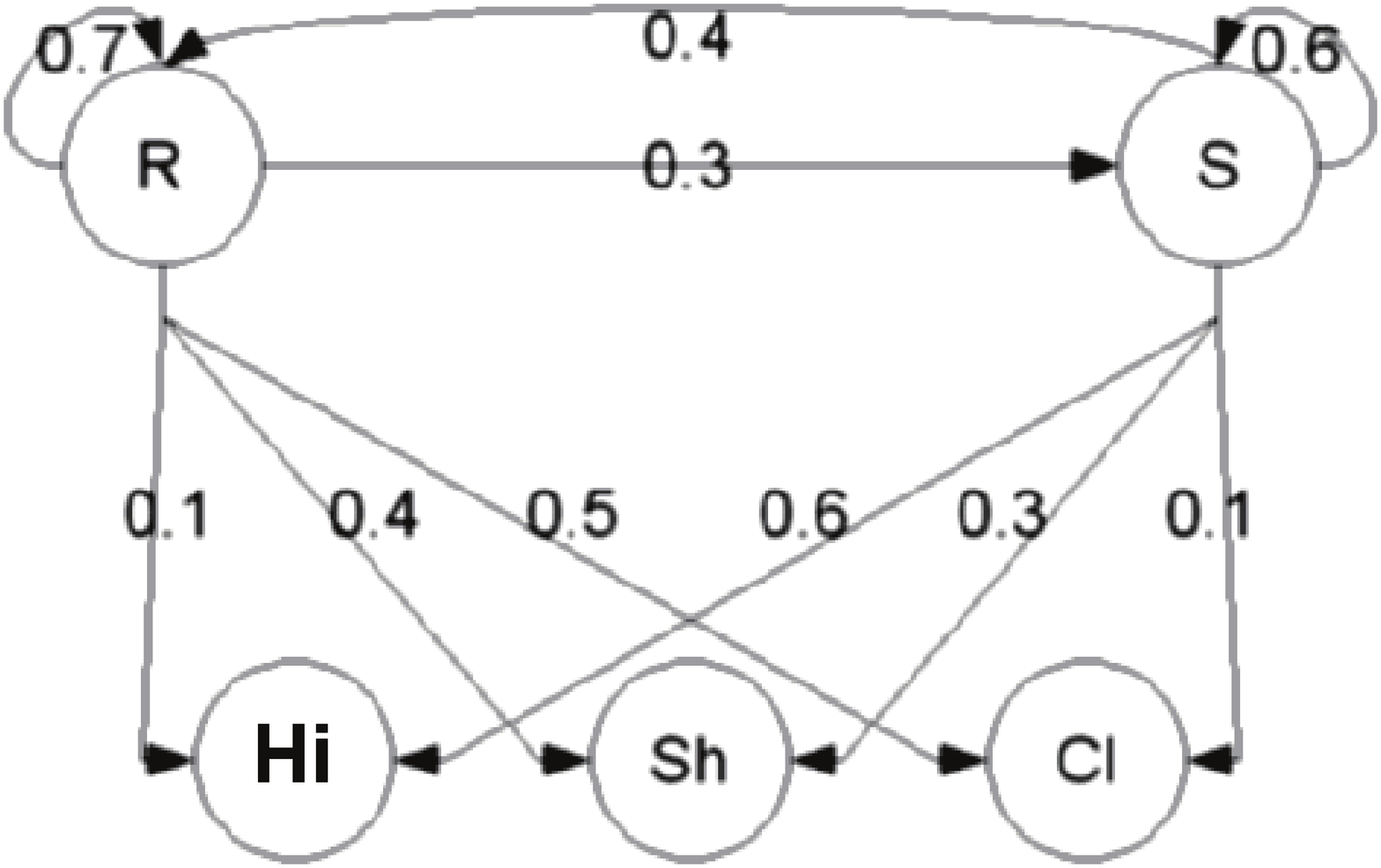}
  \caption{Hidden Markov Model for the weather and activities example}\label{Hidden-Markov-Model-for}
\end{figure}

Let us assume that some person is located in a far away city and has three activities. He carries out only one of these activities per day and then he tells us about it. The activities are hiking, shopping, and cleaning. Let us assume further that the selected activity for a day depends on the state of the weather that day. It is quite clear that the weather state every day cannot be specified deterministically. We can, however, set probabilities for the weather being rainy or sunny. After that, we try, by knowing the activity the person did, to predict the weather state that accompanied that activity. In this example, the weather state is a Hidden Markov Series and it has two states: rainy or sunny. We call this series "Hidden" because we do not know its state directly. Instead, we monitor other events to help us predict the weather state. These events are the activities of the person which are hiking, shopping, and cleaning. On the other hand, we are certain of the activity that was carried out, hence, we call the activities "observations". This whole system is called a Hidden Markov Model. Let us set up a mathematical definition for this system:
\begin{equation*}
    \begin{array}{l}
    \mathrm{States={Rainy,Sunny}}\\
    \mathrm{Observations={Hike,Shop,Clean}}\\
    \mathrm{Portability\; Set\;\#1:}\\
    \mathrm{P(Rainy)=0.6,\;P(Sunny)=0.4}\\
    \mathrm{Portability\; Set\;\#2:}\\
    \mathrm{P(Rainy\rightarrow Rainy)=0.7,\;P(Rainy\rightarrow Sunny)=0.3}\\
    \mathrm{P(Sunny\rightarrow Rainy)=0.4,\;P(Sunny\rightarrow Sunny)=0.6}\\
    \mathrm{Portability\; Set\;\#3:}\\
    \mathrm{P(Rainy\rightarrow Hike)=0.1,\;P(Rainy\rightarrow Shop)=0.4,}\\
    \mathrm{P(Rainy\rightarrow Clean)=0.5}\\
    \mathrm{P(Sunny\rightarrow Hike)=0.6,\;P(Sunny\rightarrow Shop)=0.3,}\\
    \mathrm{P(Sunny\rightarrow Clean)=0.1}
    \end{array}
\end{equation*}
The first set of probabilities represents the initial state of the system, and the second represents the probabilities of system transition from one state to another, while the third set represents the probabilities that a certain weather state accompanies a certain person activity. To solve the problem posed by this example, we have to discuss the Viterbi Algorithm. Figure \ref{Hidden-Markov-Model-for} illustrates the Hidden Markov Model for this example.

\subsubsection{The Viterbi Algorithm}\label{The-Viterbi-Algorithm}
The purpose of this algorithm is to determine the highest probability series of hidden states that generated a specific output in a Hidden Markov System \cite{Viterbi1967}. To apply this algorithm, the following assumptions must hold:
\begin{enumerate}
    \item All observations and hidden events happen in successive series: This algorithm assumes that the system we are dealing with is a state machine, and it has a group of a limited number of states (although the group may be large), and that at any time the system has one deterministically specified state.
    \item Every observation must correspond to one hidden event and vice versa: This algorithm relies on the fact that transition between states is subject to some measure such as time. Thus, observations and hidden states happen in pairs so that every observation correspond to a state and vice versa.
    \item The calculation of the series of hidden state at some moment $t$ must depend only on the events and observations that happened at that moment $t$ and on the calculated probabilistic series for the previous moment $t-1$.
\end{enumerate}
To make the Viterbi Algorithm clearer, we will use the following notation:
\begin{itemize}
    \item $sp$: The State Probabilities (the first set of probabilities in the previous example).
    \item $tp$: The Transition Probabilities (the second set of probabilities in the previous example).
    \item $ep$: The Emission Probabilities (the third set of probabilities in the previous example).
    \item $y$: The Sequence of Observations (the output sequence that was observed).
\end{itemize}
Applying this algorithm to our previous example, we find that the probability of seeing the observations (hiking, shopping, and cleaning) is $0.033612$, and that the corresponding hidden path is (sunny, rainy, rainy, rainy). This path includes $4$ states because the third observation appeared as a result of transition from the third state to the fourth one.

\subsection{Applications of Hidden Markov Models}\label{Applications-of-Hidden-Markov-Models}
Hidden Markov Models have many applications in the fields of Speech Recognition, Optical Character Recognition, and Bioinformatics. However; we are concerned with applying them to recognize natural languages. We can view the grammars of a natural language as hidden states of a system, and the words and sentences generated using these grammars as observations. This way we can specify, through applying the Viterbi Algorithm, the grammars using which a sentence was formed, and thus specify whether this sentence is a natural language sentence or not. We discuss two examples on Hidden Markov Models that generate a subgroup of the group of grammatically correct sentences in the Arabic language:\\
Sample One:
\begin{itemize}
    \item State $1$: \RL{asm, .dmyr mrfw`}
    \item State $2$: \RL{asm, .sfT, .zrf, ^gAr wm^grwr}
    \item State $3$: \RL{.sfT, .zrf, ^gAr wm^grwr}
\end{itemize}
This sample generates sentences known as \RL{^gmlaT asmyT} (Arabic for noun sentences) in the Arabic language.
Sample Two:\\
\begin{itemize}
    \item State $1$: \RL{f`l}
    \item State $2$: \RL{asm, .dmyr mrfw`, lA ^say'}
    \item State $3$: \RL{asm, .zrf, ^gAr wm^grwr}
\end{itemize}
This sample generates sentences known as \RL{^gmlaT f`lyT} (Arabic for verb sentences) in the Arabic Language.

\subsubsection{Using Hidden Markov Models to Recognize Arabic}\label{Using-Hidden-Markov-Models}
The first step in recognizing sentences of a natural language such as Arabic is to construct a special Markov Model that is capable of generating all grammatically correct sentences. Then we can apply the Viterbi Algorithm to calculate the probability that a sentence formed of Arabic words really belongs to the Arabic language. This can be achieved by following these steps:
\begin{enumerate}
    \item Build a series of observations using the words of the sentence that you want to calculate its probability of being a grammatically correct sentence by categorizing words according to their types (noun, verb, adjective, ...).
    \item Apply the Viterbi Algorithm with these input parameters:
    \begin{itemize}
        \item $y$: The list that was built in step $1$.
        \item $x$: A list of all word types in the Arabic language.
        \item $sp$: We get the value of this parameter from the Markov Model of the Arabic language.
        \item $tp$: We get the value of this parameter from the Markov Model of the Arabic language.
        \item $ep$: We get the value of this parameter from the Markov Model of the Arabic language.
    \end{itemize}
\end{enumerate}
After calculating the probability, we compare it with a certain threshold to determine whether the sentence is accepted or not. Of course, having a large number of sentences allows us to increase the accuracy of the obtained results.

\section{Grammar Substitution}\label{Grammar-Substitution}
\begin{figure}
  \centering
  \includegraphics[scale=0.15]{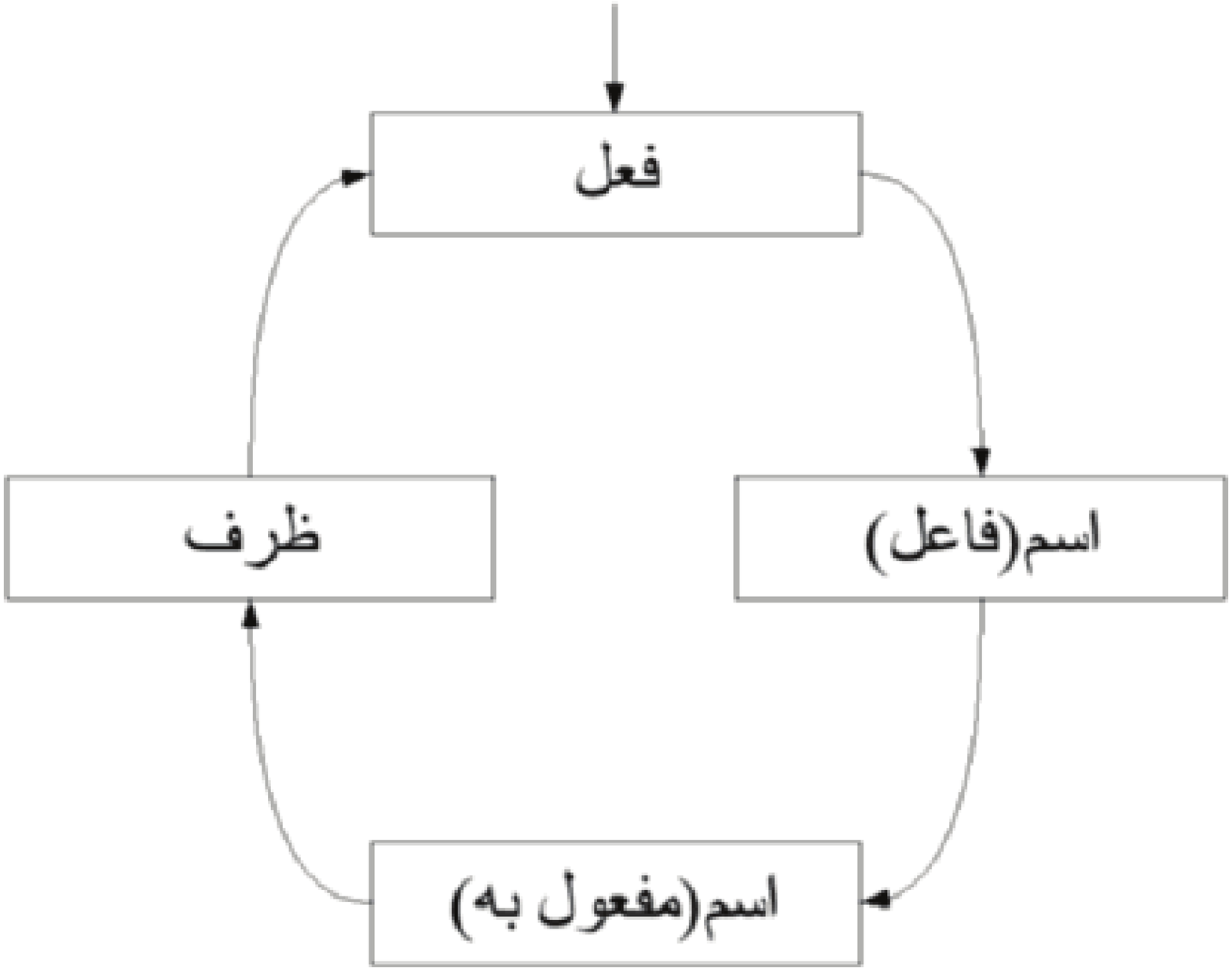}
  \caption{The Hidden Markov Model used to generate grammatically correct Arabic sentences}\label{The-Hidden-Markov-Model-used}
\end{figure}

We can now set up a method for hiding encrypted data in innocent natural language sentences, thereby fooling observer-supporting software into thinking that the encrypted data is not encrypted at all.
\resizebox{1.0\hsize}{!}{Our suggested method will be based on the Word Substitution method}, but instead of selecting a word from a dictionary using its number only, we will satisfy additional criteria that guarantee that the final result is going to be a natural language text. The first step is to set up the grammars using which the text will be formed. These grammars must be part of the Arabic language grammars. Actually, there are certain patterns that happen frequently in the Arabic language. Let us assume the following two patterns:
\begin{itemize}
    \item \RL{f`l, asm (fA`l), asm (mf`wl bh), .zrf}
    \item \RL{asm (mbtda'a), asm (_hbr), .hrf ^gr, asm (m^grwr)}
\end{itemize}
We start the operation of encrypted data hiding using the Word Substitution method, but this time we select words of certain types in a way that adheres to the first grammar, for example. That means, we divide our dictionary into three parts: a part for words of type \RL{f`l} (Arabic for verb), a part for words of type \RL{asm} (Arabic for noun), and finally a part for words of type \RL{.zrf} (Arabic for adverb). During the hiding operation, we select from the dictionary of \RL{f`l} first, then from the dictionary of \RL{asm}, then from the dictionary of \RL{asm}, and finally from the dictionary of \RL{.zrf}. This way we get a grammatically correct sentence and, at the same time, one that has the characteristics of the Word substitution method. The reverse operation happens the same way using the dictionary corresponding to the type of the current word. Figure \ref{The-Hidden-Markov-Model-used} illustrates the Hidden Markov Model used in the previous example to generate grammatically correct Arabic sentences.

\subsection{Grammatical Substitution Algorithm}\label{Grammatical-Substitution-Algorithm}
We build a dictionary of words categorized according to their types. The larger the dictionary, the more efficient the output of the algorithm. Then we follow these steps:
\begin{enumerate}
    \item Get the type of the next word from the grammar and determine the number of elements $m$ in the dictionary corresponding to this type.
    \item Calculate the number of bits that should be read from the encrypted data using the formula:
    \begin{equation*}
        n=\mathit{floor}(\log_{2}m)
    \end{equation*}
    \item Read the next $n$ bits from the encrypted data, convert them to an integer, and then select the word with the corresponding number from the dictionary.
    \item Return to step $1$ until there is no more encrypted data.
\end{enumerate}
For example, when we hide the following random bytes:
\begin{equation*}
    \mathrm{d9\;e6\;59\;42}
\end{equation*}
We may get the following Arabic sentence:
\begin{equation*}
    \RL{st_hfy 'ashm 'a_d_A}
\end{equation*}
Despite the fact that this Arabic sentence has no meaning, it is grammatically correct and enough to bypass observer-supporting software as an innocent text.

\subsection{Reversal of Grammatical Substitution Algorithm}\label{Reversal-of-Grammatical-Substitution-Algorithm}
To get the encrypted data back again from the natural text, we must use the same dictionaries and grammars that were used in the hiding operation. Then we follow these steps:
\begin{enumerate}
    \item Read a word from the natural text, and get the next type from the grammar.
    \item Calculate the number of bits that should be written as encrypted data using the formula:
    \begin{equation*}
        n=\mathit{floor}(\log_{2}m)
    \end{equation*}
    where $m$ is the number of elements in the dictionary corresponding to the type.
    \item Determine the number of the word in the dictionary, convert it to binary representation, clear bits whose order is higher than $n$, and then take the rest of the bits.
    \item Returning to step $1$ until there is no more natural text words.
\end{enumerate}

\begin{figure}
  \centering
  \includegraphics[scale=0.28]{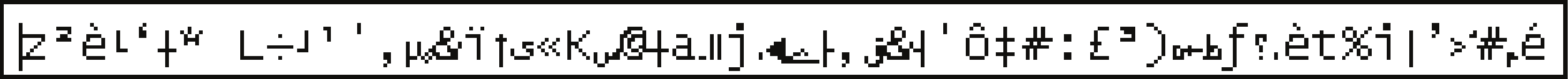}
  \caption{The experiment text encrypted with AES ($256$ bits key)}\label{The-experiment-text-encrypted-with-AES}
\end{figure}

\subsection{Grammatical Substitution Redundancy}\label{Grammatical-Substitution-Redundancy}
As with Word Substitution, The Redundant Data Percentage in Grammatical Substitution is far less than it is in Sentence Substitution. However; Grammatical Substitution has the advantage of valid grammatical structure when compared to Word Substitution. The Redundant Data Percentage depends on the size of the dictionaries. The more the words in these dictionaries, the more the bits that can be encoded into one word. Contrary to Word Substitution that uses one dictionary for all substitutions, Grammatical Substitution uses a separate dictionary for every word type, therefore, we need larger dictionaries in Grammatical Substitution to achieve the same redundancy in Word Substitution.

\subsection{Algorithm Implementation}\label{Algorithm-Implementation}
Our system of hiding encrypted data relies heavily on having large dictionaries containing words categorized according to their types (noun, verb, adjective, ...), and on having grammars that form a subgroup of the syntactic grammars of the target language. Having that at our disposal, we can convert the random bits of encrypted data into natural language sentences. We discovered that the quality of the output enhances by increasing the number of dictionaries used and the complexity of the grammars. Unfortunately, the quality of the Arabic dictionaries available for academic use were not satisfying at all. And the best dictionary we were able to obtain was formed of $81011$ words only; this is nothing compared to the $4$ million words in the paper dictionary \RL{lsAn al`rb} (Arabic for Arab Tongue). The problem of quality and availability of Arabic dictionaries can be overcome by designing a flexible hiding system that allows users to update the dictionary and the grammars. Let us now calculate the Redundant Data Percentage when using the dictionary mentioned above. We know that the formula that relates the size of the dictionary $m$ to the number of bits $n$ in Word Substitution is:
\begin{equation*}
    n=\mathit{floor}(\log_{2}m)
\end{equation*}
Substituting $m$ for $81011$, we find that $n=16$ bits; means we can encode every two bytes in one word. The average length of a word in this dictionary is around $4.2$ letters. If we assumed that an Arabic letter needs around $2$ bytes of storage space, we arrive at the following average length of a dictionary word:
\begin{equation*}
    \mathit{length}=4.2\cdot2\cdot8=67.2\mathit{\; bits}
\end{equation*}
And the Redundant Data Percentage would be:
\begin{equation*}
    r=(\mathit{length}-n)/\mathit{length}=76.2\%
\end{equation*}
Note that the percentage increases in Grammatical Substitution method; however, it depends on the grammars used. And as we said earlier, this percentage can be decreased by using larger dictionaries.

\section{Testing the Cryptosystem}\label{Testing-the-Cryptosystem}
\begin{figure}
  \centering
  \includegraphics[scale=0.28]{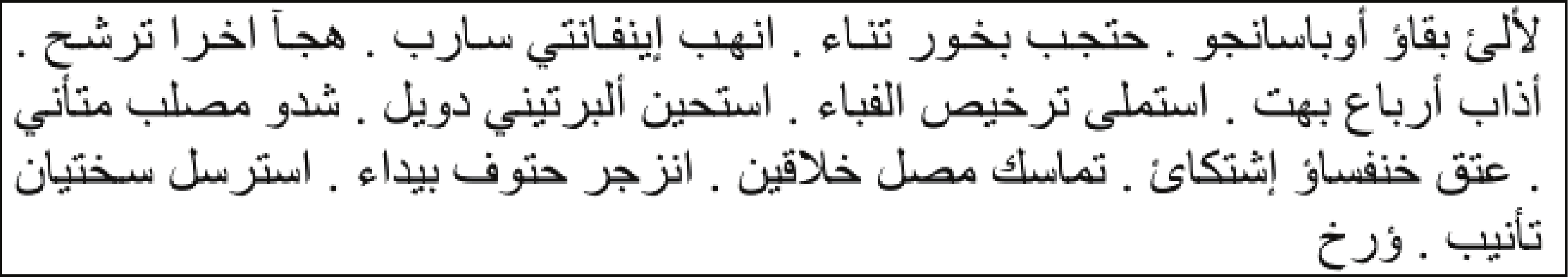}
  \caption{The innocent Arabic string corresponding to the encrypted experiment text shown in Figure \ref{The-experiment-text-encrypted-with-AES}}\label{The-innocent-Arabic-string}
\end{figure}

To test our cryptosystem, we developed a computer program in the Java Programming Language. We had the message "Do not attribute to malice what can be explained by stupidity!" encrypted with AES ($256$ bits key). The encrypted text was as shown in Figure \ref{The-experiment-text-encrypted-with-AES}. We loaded the encrypted text into our program. And the program converted it into the following hexadecimal string:
\begin{equation*}
    \begin{array}{c}
    \mathrm{7A\;B2\;E8\;03\;91\;10\;2A\;20\;4C\;F7\;04\;B9\;27\;2C\;B5\;F2\;26\;EF}\\
    \mathrm{18\;EC\;AB\;4B\;D3\;40\;0A\;10\;61\;DC\;13\;6A\;A1\;11\;DE\;2C\;19}\\
    \mathrm{FF\;26\;17\;27\;F4\;87\;23\;3A\;A3\;C0\;28\;B3\;1B\;D8\;83\;BF\;A1\;E8}\\
    \mathrm{74\;25\;69\;05\;92\;3E\;F5\;23\;E3\;1E\;E9}
    \end{array}
\end{equation*}
To test the Hush encryption, we applied our cryptosystem to the previous hexadecimal string to obtain the innocent Arabic string shown in Figure \ref{The-innocent-Arabic-string}. To test the Hush decryption, we applied our cryptosystem again to the previous Arabic string to obtain the same hexadecimal string, listed above, that represents the original AES-encrypted message.

\section{Results of Applying the Statistical Batteries}\label{Results-of-Applying-the-Statistical-Batteries}
We have analyzed The Hush Cryptosystem using the statistical battery ENT. Random Front End \cite{Hussein}, an open-source project we developed around the end of $2008$, eases the application of this battery. The results corresponding to the ENT Battery for a plain text file of around $29$ kilobytes are shown in Table \ref{Results-of-applying}. Note how close the results for plain text (1st column) to those for Hush-encrypted text (3rd column), and how far they both are from the results for AES-encrypted text (2nd column). The Randomness Degree of encrypted data is decreased, and observer-supporting software is fooled, and will not be able to tell plain from encrypted data by employing statistical tests.

\begin{table}
    \centering
    \caption{Results of applying the ENT battery to a plain text file of around $29$ kilobytes}
    \label{Results-of-applying}
    \begin{tabular}{llll}
    \toprule
    Test & Plain text & AES- & Hush- \\
    & & encrypted & encrypted \\
    \midrule
    Entropy & $4.916529$ & $7.995169$ & $4.412139$ \\
    & bits/byte & bits/byte & bits/byte \\

    Optimum & $38\%$ & $0\%$ & $44\%$ \\
    Compression & & & \\

    Chi Square & $418649.79$ & $196.94$ & $1965293.88$ \\
    Distribution & & & \\

    Arithmetic & $84.1431$ & $127.1892$ & $164.2434$ \\
    Mean & & & \\

    Monte Carlo & $3.98036$ & $3.16073$ & $2.15373$ \\
    Value For Pi & (error & (error & (error \\
    & $26.70\%$) & $0.61\%$) & $31.44\%$) \\

    Serial Correlation & $0.287639$ & $0.007207$ & $0.163473$ \\
    Coefficient & & & \\
    \bottomrule
    \end{tabular}
\end{table}

\section{English, Why Not?}\label{English-Why-Not}
Let us employ English rather than Arabic for the implementation of our cryptosystem. We use different grammars for Hidden Markov Models, and different dictionaries to generate grammatically correct English sentences. The approach is remarkably the same. However; the considerable advancements in the field of Natural Language Understanding (NLU), particularly the Cyc NL subsystem \cite{Cycorpa} of the controversial Cyc project with its large English knowledge base poses many difficulties. An observer-supporting software linked to OpenCyc \cite{Cycorp}, and utilizing its concepts and facts pertaining to various realms of knowledge is able to detect spurious sentences encrypted with our cryptosystem and then provide them to the observer; without having to rely on statistical tests. \resizebox{1.0\hsize}{!}{Our cryptosystem is not a Natural Language Generator, and this}\balancecolumns is not a combat between generation and understanding, otherwise, the cryptosystem is never handy nor portable. On the other hand, it is highly unlikely for the English knowledge base of the Cyc project to be ported or provided in other natural languages. Therefore, using Arabic for the implementation of our cryptosystem is a judicious decision.

\section{Acknowledgments}
I am deeply grateful for the proactive and valuable participation of Jonna Stugufors who made many useful suggestions that improved this paper.

\bibliographystyle{abbrv}
\bibliography{hush}

\end{document}